\documentclass[aps,prl,bibnotes,twocolumn,showpacs,preprintnumbers,amsmath,amssymb,superscriptaddress,floatfix]{revtex4-2}
\usepackage[T1]{fontenc}
\usepackage[ansinew]{inputenc}
\usepackage{graphics}
\usepackage{dcolumn}
\usepackage{soul}
\usepackage{bm}
\usepackage{amsthm}
\usepackage{amsmath}
\usepackage{amssymb}
\usepackage{diagbox}
\usepackage{hyperref}
\usepackage{url}
\usepackage{xcolor}

\def\afflux{Department of Physics and Materials Science, University of Luxembourg, 162A~Avenue de la Faiencerie, L-1511 Luxembourg, Grand Duchy of Luxembourg}
\def\affias{Institute for Advanced Studies, University of Luxembourg, Campus Belval, L-4365 Esch-sur-Alzette, Grand Duchy of Luxembourg}
\def\affill{Institut Laue-Langevin, 71~Avenue des Martyrs, 38042~Grenoble, France}
\def\affghent{Department of Solid State Sciences, Ghent University, Krijgslaan 285/S1, 9000 Ghent, Belgium}
\def\affpuerto{Departamento de Nanociencia y Nanotecnolog{\'i}a, Instituto de Ciencia de Materiales de Madrid, ICMM/CSIC, C/Sor Juana In{\'e}s de la Cruz~3, 28049~Madrid, Spain}
\def\affdavidegenova{Department of Chemistry and Industrial Chemistry \& INSTM RU, University of Genoa, Via Dodecaneso~31, 16146~Genova, Italy}
\def\affdavideroma{Institute of Structure of Matter, National Research Council, nM2-Lab, Via Salaria km 29.300, Monterotondo Scalo, Roma, Italy}
\def\affluis{Department CITIMAC, Facultad de Ciencias, Universidad de Cantabria, 39005 Santander, Spain}
\def\affmpg{Max Planck Institute for the Structure and Dynamics of Matter, Luruper Chaussee~149, D-22761 Hamburg, Germany}
\def\affcfel{Center for Free-Electron Laser Science, Notkestra\ss e~85, D-22607 Hamburg, Germany}

\begin{document}

\title{Polarized neutron scattering as a probe for vortex-type spin correlations in iron oxide multicore assemblies}

\author{Venus Rai}\email{venus.rai@uni.lu}
\address{\afflux}
\author{Ivan Titov}
\address{\afflux}
\author{Elizabeth M.\ Jefremovas}
\address{\afflux}
\address{\affias}
\author{\v{S}tefan Li\v{s}\v{c}\'{a}k}
\address{\afflux}
\author{Sivarenjini Shan}
\address{\afflux}
\address{\affill}
\author{Nina-Juliane Steinke}
\affiliation{\affill}
\author{Jonathan Leliaert}
\address{\affghent}
\author{\'Alvaro Gallo-C\'ordova}
\address{\affpuerto}
\author{Mar{\'i}a P. Morales}
\address{\affpuerto}
\author{Davide Peddis}
\author{Pierfrancesco Maltoni}
\address{\affdavidegenova}
\address{\affdavideroma}
\author{Luis Fern{\'a}ndez Barqu{\'i}n}
\address{\affluis}
\author{Andreas Michels}\email{andreas.michels@uni.lu}
\address{\afflux}
\author{Michael P.\ Adams}\email{michael.adams@mpsd.mpg.de}
\address{\affmpg}
\address{\affcfel}

\date{\today}

\begin{abstract}
We report an experimental investigation of the magnetic microstructure of iron oxide multicore assemblies by means of polarized small-angle neutron scattering (SANS). Guided by a recently developed analytical theory for vortex-state magnetic nanoparticles, we provide a quantitative comparison between the measured and calculated cross sections, revealing signatures that are consistent with vortex-type magnetization configurations at low applied magnetic fields. In particular, the field evolution and the characteristic isotropic ring-type feature of the spin-flip scattering intensity at intermediate momentum transfers are in line with the formation of flux-closure states. The latter are stabilized by the interplay of exchange, Zeeman, and magnetostatic energies. The methodology allows for a statistically significant characterization of vortex states in densely packed nanoparticle systems, thereby complementing surface-sensitive techniques that are commonly limited to the observation of spin structures in individual particles.
\end{abstract}

\date{\today}

\maketitle


{\it Introduction}---The equilibrium spin configuration of magnetic nanoparticles remains a central question in the field of nanomagnetism. For sufficiently small particles, the exchange interaction dominates the magnetostatic self-energy and stabilizes a single-domain state whose quasistatic hysteretic response is well described by the Stoner-Wohlfarth model~\cite{sw48}. With increasing particle size, however, the balance between micromagnetic energy contributions changes: the growing influence of the magnetostatic interaction promotes flux-closure arrangements and can stabilize nonuniform spin textures, including curling and vortex states. Such behavior was already anticipated in the seminal work of Brown~\cite{brown} and later incorporated into the general micromagnetic framework developed by Aharoni~\cite{aharonibook}.

Beyond their fundamental relevance, magnetic nanoparticles play a key functional role in applications that rely on field-induced energy dissipation, including magnetic hyperthermia, environmental remediation, or highly efficient catalysis~\cite{gavilan2025magnetic}. Monte Carlo and micromagnetic simulations predict that nanoparticles with a significant degree of spin disorder may exhibit higher specific absorption rates compared to their ordered counterparts~\cite{lappas2019,eliss2026}. In particular, vortexlike spin configurations have been shown to reach performance metrics similar to those of single-domain particles (e.g., \cite{Usov2018_VortexHyperthermia,eliss2026,jefremovas2026nanoscale}). Moreover, an inherent advantage of vortex-type textures is their small but finite net magnetization, which implies a reduced stray field that prevents particle agglomeration, detrimental to heating performance~\cite{etheridge2014accounting}.

Despite these considerations, the macrospin approximation, where the magnetization within each particle is treated as uniform and homogeneous~\cite{nowakprb2005}, remains a standard framework for interpreting the magnetic behavior of fine-particle systems. Its widespread and continued use reflects not only its conceptual simplicity, but also practical limitations: experimental techniques capable of resolving spin configurations inside nanoparticles are still relatively scarce. To date, vortexlike spin textures in nanoparticles and nanostructures have been observed primarily through surface-sensitive methods, including magnetic force microscopy and Lorentz transmission electron microscopy (e.g., \cite{gatel2015,vortexexp0,vortexexp2,Niraula2023_VortexNanospheres}). More recently, x-ray transmission microscopy has allowed the investigation of vortex states with a greater penetration depth, as demonstrated in individual nanoflowers~\cite{BATLLE2024}.

Although these experimental techniques yield detailed, particle-specific insights, they are inherently limited when it comes to probing large ensembles. Achieving a statistically significant characterization of vortex states in densely packed systems requires techniques that are capable of interrogating a substantial sample volume and probe many particles simultaneously. This challenge is compounded by the intrinsic disorder in such assemblies: the magnetic easy axes of the particles are typically randomly oriented, and at the low applied magnetic fields where vortices may emerge, an additional distribution of vortex axes must also be considered.

In this work, we introduce a neutron-scattering approach for the quantitative analysis of vortexlike magnetic correlations in bulk ensembles of magnetic nanoparticles. We employ polarized small-angle neutron scattering (SANS), which is uniquely suited for resolving spin structures on mesoscopic length scales and within the volume of magnetic materials~\cite{rmp2019,michelsbook}. Building on a recently developed theory of small-angle scattering from vortices~\cite{adamsprb2024no2,adamsprr2026}, we analyze the spin-flip SANS cross section of an iron oxide nanoflower sample and obtain model-based evidence for vortex-type spin correlations. We refer to the Supplemental Material~\cite{nanoflower2026sm} for additional information and results supporting this study (see also Refs.~\cite{michels2010epjb,kryckaprl2014,zakutna2020,ukleev2024,squires,dewhurst2016,illdoi-nanoflower,wildes06,dewhurst2023,kinning1984,mumax3new,laura2020,evelynprb2024,adamsjac2026} therein).




\begin{figure}[tb!]
\centering
\resizebox{1.0\columnwidth}{!}{\includegraphics{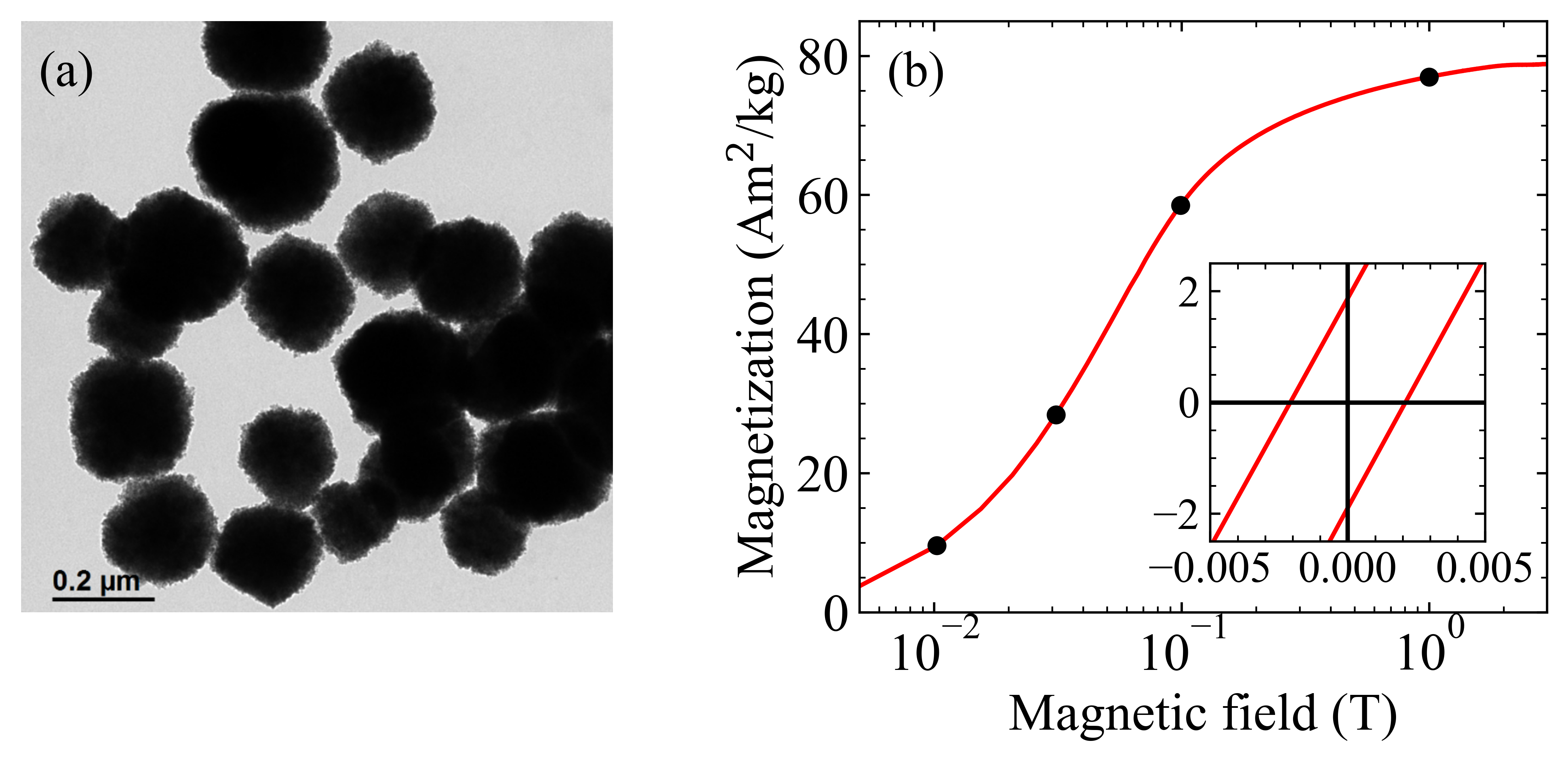}}
\caption{(a)~Transmission electron microscopy image of iron oxide nanoflowers with an average flower size of $D = 220 \pm 20 \, \mathrm{nm}$. Each nanoflower consists of nanocrystallites with an average size of about $10 \, \mathrm{nm}$ determined from the analysis of wide-angle x-ray diffraction data~\cite{nanoflower2026sm}. (b)~Room-temperature magnetization data of the $220 \, \mathrm{nm}$ nanoflower sample (log-linear scale, only the upper right quadrant is shown). The black dots mark the field values of the SANS measurements. The inset highlights the hysteretic behavior at small fields.}
\label{fig1}
\end{figure}

{\it Experimental}---The neutron experiment was conducted at the D33 instrument at the Institut Laue-Langevin (ILL), Grenoble, France (see \cite{nanoflower2026sm} for further details).
The samples under study are iron oxide nanoflowers, which were prepared by wet-chemical processing (see, e.g., Ref.~\cite{BATLLE2024} for details regarding sample synthesis and structural and magnetic characterization). Figure~\ref{fig1}(a) shows an electron microscopy image of an iron oxide nanoflower sample that has been investigated in this study. The shape of the nanoflowers is approximately spherical and we find an average nanoflower diameter of about $220 \, \mathrm{nm}$. Each aggregate is composed of several nanocrystallites (so-called cores) of about $10 \, \mathrm{nm}$ in size. The magnetization curve [Fig.~\ref{fig1}(b)] indicates that the approach-to-saturation regime is reached for applied fields larger than about $1 \, \mathrm{T}$.


{\it Linear vortex model for the spin-flip SANS}---In Ref.~\cite{adamsprb2024no2}, a magnetization model for a spherical nanoparticle that contains a vortex-type spin texture has been introduced. The magnetization of the particle with radius $R$ consists of a uniform part of magnitude $m_0$ and a linear vortex term of magnitude $m_1$. More specifically, the local magnetization vector field is written as:
\begin{align}
\mathbf{M}(\mathbf{r}) = m_0 \mathbf{e}_z + m_1 \mathbf{v}(\mathbf{r}) ,
\label{eq:LinearVortexModel}
\end{align}
where $\mathbf{e}_z = \{ 0, 0, 1 \}$ is the unit vector in the $z$~direction, $\mathbf{v}(\mathbf{r}) = \{ -y, x, 0 \}$ is the linear vortex field, and $\mathbf{r} = \{ x,y,z \}$ is the position vector with reference to the local vortex frame. This Ansatz results in the following expression for the 2D spin-flip SANS cross section $d\Sigma_{\mathrm{sf}} / d\Omega$ of a dilute ensemble of randomly-oriented spherical vortex-state nanoparticles:
\begin{widetext}
\begin{align}
\label{eq:SpinFlip2DSANScrossSection}
\frac{d\Sigma_{\mathrm{sf}}}{d\Omega}(q, \theta) &= \frac{W}{8}
\left[ m_0 f(qR) \right]^2 \times \left[ 12 - (\cos^2\alpha_{\mathrm{c}} + \cos\alpha_{\mathrm{c}})(3\cos^2(2\theta) + 2\cos(2\theta) + 3) + 4 \cos(2\theta) \right] \\
& + \frac{W}{2} \left[ m_1 R f'(qR) \right]^2 \times \left[ 3 - (2 \cos^2\alpha_{\mathrm{c}} + 2 \cos\alpha_{\mathrm{c}} - 1)\cos(2\theta) \right] \nonumber ,
\end{align}
\end{widetext}
where $W$ is a scaling constant, $f(qR) = \frac{j_1(qR)}{qR}$ with $j_1(u)$ being the first-order spherical Bessel function, $f'(qR) = \frac{df(qR)}{d(qR)}$, and $\theta$ is the angle between the scattering vector $\mathbf{q}$ and the applied magnetic field $\mathbf{H}_0 \parallel \mathbf{e}_z$ (see Fig.~1 in \cite{nanoflower2026sm}). The quantity $0^{\circ} \leq \alpha_{\mathrm{c}} \leq 90^{\circ}$ is a field-dependent conical opening angle describing the orientation of the vortex rotation axis relative to $\mathbf{H}_0$. Note that $\alpha_{\mathrm{c}} = 0^{\circ}$ in the saturated case, and that $\alpha_{\mathrm{c}}$ increases with decreasing field. Equation~(\ref{eq:SpinFlip2DSANScrossSection}) is expected to be valid in the low-field regime when vortices might have nucleated. When plotted as a function of the ratio $m_1 R/m_0$ and $\alpha_{\mathrm{c}}$, the 2D $\frac{d\Sigma_{\mathrm{sf}}}{d\Omega}(q, \theta)$ exhibits a variety of angular anisotropies~\cite{adamsprr2026}.

Averaging Eq.~(\ref{eq:SpinFlip2DSANScrossSection}) over the detector angle $\theta$, i.e., $(2\pi)^{-1} \int_0^{2\pi} (...) d\theta$, yields the 1D quantity~\cite{adamsprb2024no2}
\begin{align}
\label{eq:SpinFlip1DSANScrossSection}
I_{\mathrm{sf}}(q) &= \frac{3W}{16} \left[ m_0 f(qR) \right]^2 \left( 8 - 3 \cos^2\alpha_{\mathrm{c}} - 3 \cos\alpha_{\mathrm{c}} \right) \\
&+ \frac{3W}{2} \left[ m_1 R f'(qR) \right]^2 \nonumber .
\end{align}
Equations~(\ref{eq:SpinFlip2DSANScrossSection}) and (\ref{eq:SpinFlip1DSANScrossSection}), which straightforwardly follow from the magnetization model Eq.~(\ref{eq:LinearVortexModel}), describe the spin-flip SANS of a dilute system of vortex-state nanoparticles, i.e., they do not account for the effect of dense packing (interparticle interference) and for a distribution of particle sizes~\cite{borchers2025}.

Experimental neutron data contain, to a certain extent, the above effects, i.e., dense packing and polydispersity. Therefore, to quantitatively describe the real measured system, we have used the following phenomenological expression:
\begin{align}
\langle I_{\mathrm{sf}}(q) \rangle = S(q, \overline{R}, \eta) \overline{I_{\mathrm{sf}}}(q, \overline{R}, \sigma) ,
\label{eq:finalmodel}
\end{align}
where $S(q, \overline{R}, \eta)$ denotes the analytically-known Percus-Yevick hard-sphere structure factor (with a particle volume fraction of $\eta$)~\cite{hansenbook1986,chentartagliabook},
and
\begin{eqnarray}
\overline{I_{\mathrm{sf}}}(q, \overline{R}, \sigma) = \int_{0}^{\infty} I_{\mathrm{sf}}(q, R) f(R, \overline{R}, \sigma) dR
\label{eq:finalmodel1}
\end{eqnarray}
represents the quasi-spherical flower form factor averaged over the size distribution function $f(R, \overline{R}, \sigma)$. In our analysis, we assumed $f(R, \overline{R}, \sigma)$ to be of the lognormal type~\cite{krill98} with a median of $\overline{R}$ and a variance of $\sigma$~\cite{nanoflower2026sm}. Equation~(\ref{eq:finalmodel}) is motivated by the more familiar approach in nonmagnetic (nuclear) SANS and SAXS, where the scattering intensity of a densely packed particle system is frequently written as the product of a structure factor and a particle form factor~\cite{kotlarchyk1983,pedersen94,pedersen97}. We therefore consider the set of Eqs.~(\ref{eq:SpinFlip2DSANScrossSection})$-$(\ref{eq:finalmodel1}) as an effective model that is expected to capture the essential signatures of vortex-type spin correlations in dense nanoparticle ensembles.


\begin{figure*}[tb!]
\centering
\resizebox{1.85\columnwidth}{!}{\includegraphics{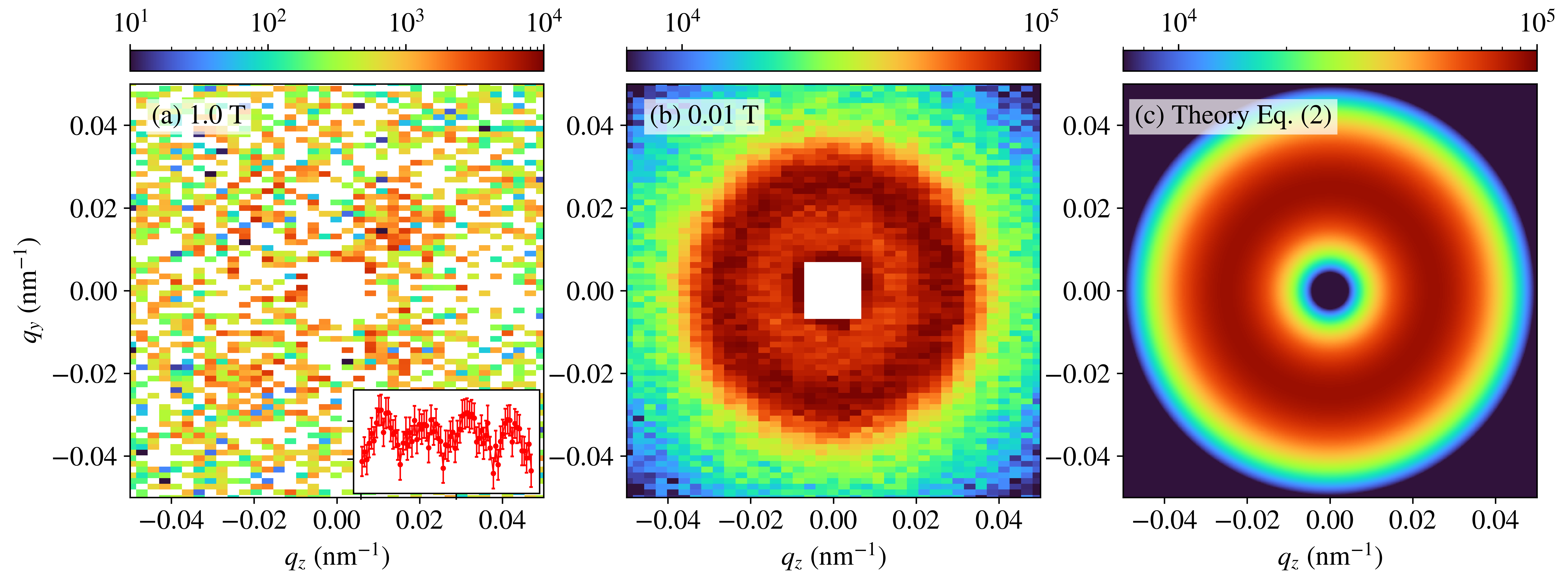}}
\caption{(a)~Two-dimensional experimental spin-flip SANS cross section $d \Sigma_{\mathrm{sf}} / d \Omega$ (in $\mathrm{cm}^{-1}$) of iron oxide nanoflowers at applied magnetic fields of (a)~$1.0 \, \mathrm{T}$ and (b)~$0.01 \, \mathrm{T}$ ($\mathbf{H}_0 \parallel \mathbf{e}_z \perp \mathbf{k}_0 \parallel \mathbf{e}_x$). The inset in (a) depicts a $\sin^2\theta \cos^2\theta$~type angular variation of $d \Sigma_{\mathrm{sf}} / d \Omega$ at $1 \, \mathrm{T}$, which is characteristic for the saturated state. (c)~Analytical prediction by the linear vortex model [Eq.~(\ref{eq:SpinFlip2DSANScrossSection})] using the following parameters: $m_1 R/m_0 = 100$, $R = 93 \, \mathrm{nm}$, and $\alpha_{\mathrm{c}} = \arccos{[(\sqrt{3}-1)/2]} = 68.5^{\circ}$~\cite{adamsprr2026}. The analytical data in (c) were scaled by a numerical factor to match the intensity range in (b). Logarithmic color scales are used in (a)$-$(c).}
\label{fig2}
\end{figure*}

{\it Results and discussion}---Before discussing the experimental neutron results, we will briefly summarize what is known concerning the microstructure of iron oxide nanoflowers. This material system, with a distinct multicore morphology, has been extensively investigated in the literature with magnetic induction heating prospects, including magnetic hyperthermia and applications beyond the biomedical field (see, e.g., Refs.~\cite{gazeau2012No1,bender2018jpcc,puerto2017,gallo2022,BATLLE2024,gavilan2025magnetic} and references therein).

Nanoflowers of the present size class ($D = 220 \, \mathrm{nm}$) cannot be strictly described as single-crystalline or polycrystalline. Rather, they are better understood as mesocrystalline structures, i.e., oriented aggregates of nanocrystals that exhibit a certain degree of crystallographic correlation across several cores. In practice, this partial ``alignment'' allows magnetic interactions between neighboring nanocrystals. Individual cores (here, around $10 \, \mathrm{nm}$ in size) can be reasonably approximated as nearly single-domain magnetic units. The magnetic behavior of the nanoflower therefore emerges from the interplay between this local single-domain character, the exchange coupling between adjacent cores sharing coherent interfaces, and the magnetostatic interaction.

With respect to their magnetic anisotropy, the magnetite/maghemite phases display a cubic magnetocrystalline anisotropy, with an additional uniaxial contribution arising from surface effects, structural disorder, and intercore interactions within the nanoflower. Micromagnetic simulations have modeled such a rich intraparticle anisotropy landscape by the combination of a grain-dependent uniaxial anisotropy direction plus a reduced intercore exchange coupling~\cite{eliss2026}, in this way allowing for an effective description that is capable of reproducing the macroscopically observed magnetic behavior of the nanoflowers~\cite{jefremovas2026nanoscale}.

\begin{figure}[tb!]
\centering
\resizebox{0.95\columnwidth}{!}{\includegraphics{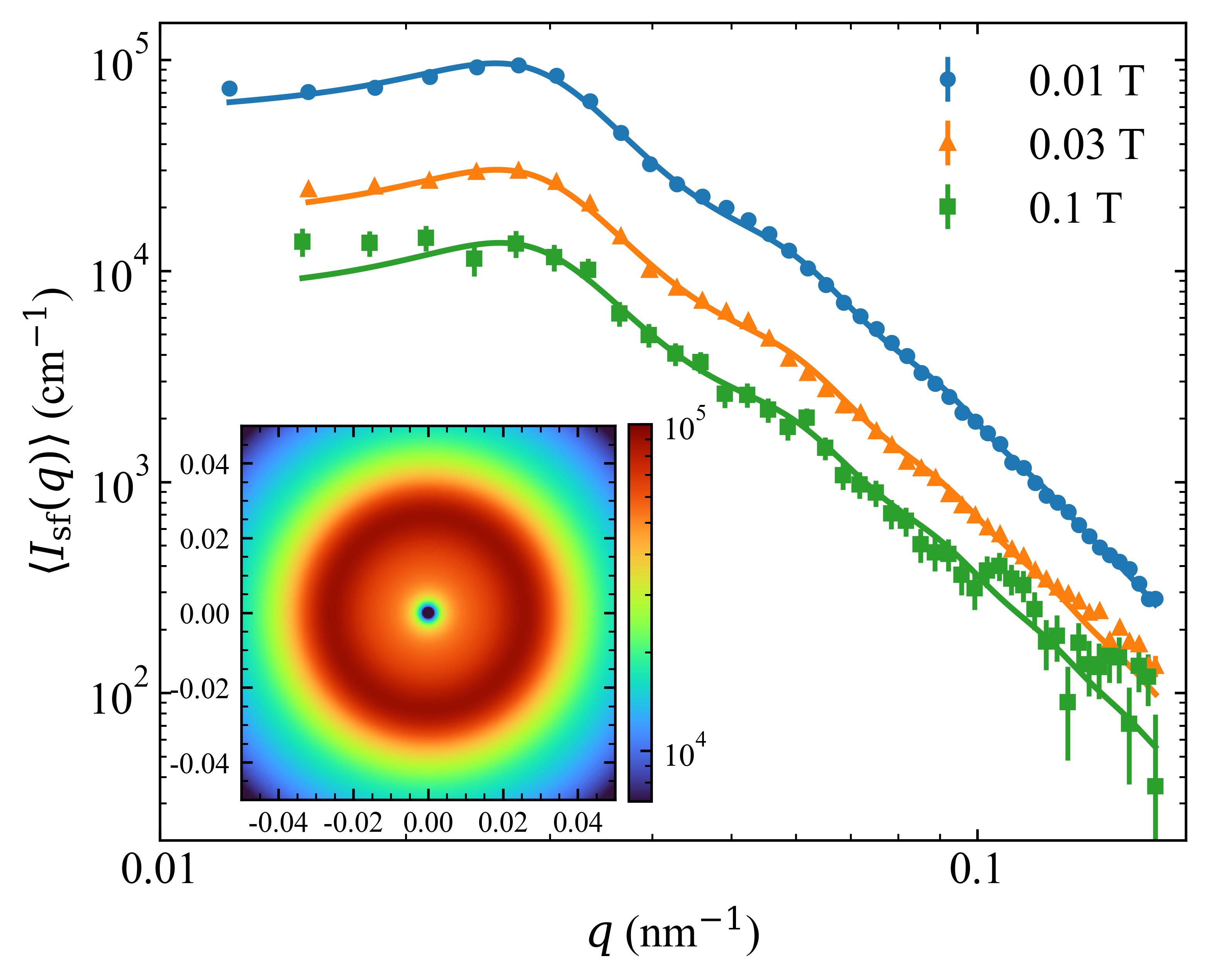}}
\caption{($\bullet$)~Experimental $\langle I_{\mathrm{sf}}(q) \rangle$ at several applied magnetic fields (see inset, log-log scale). Solid lines: fit to the linear vortex model [Eqs.~(\ref{eq:SpinFlip1DSANScrossSection})$-$(\ref{eq:finalmodel1})]. Inset: 2D spin-flip SANS at $0.01 \, \mathrm{T}$ computed based on the fit results [compare to Fig.~\ref{fig2}(b)].}
\label{fig3}
\end{figure}

Figure~\ref{fig2} displays the experimental results for the 2D spin-flip SANS cross section $d \Sigma_{\mathrm{sf}} / d \Omega$ measured on a nanoflower ensemble with a mean size of $D = 220 \, \mathrm{nm}$. We emphasize that the spin-flip SANS cross section does not contain the nuclear coherent (structural) scattering signal; $d \Sigma_{\mathrm{sf}} / d \Omega$ is dominantly due to mesoscale variations in the 3D magnetization vector field of the sample. At a field of $1 \, \mathrm{T}$ close to saturation [Fig.~\ref{fig2}(a)], the experimental $d \Sigma_{\mathrm{sf}} / d \Omega$ is very weak and exhibits the expected $\sin^2\theta \cos^2\theta$~type angular anisotropy (see inset in Fig.~\ref{fig2}(a) and compare to Eq.~(5) in \cite{nanoflower2026sm}). Reducing the field to $0.01 \, \mathrm{T}$ [Fig.~\ref{fig2}(b)] results in the appearance of an isotropic SANS pattern and in the formation of a ring structure at intermediate $q$~values~\footnote{Nanoporosity, which is conjectured to be present in this system~\cite{privcomm2026}, is likely not at the origin of the observed magnetic scattering in Figs.~\ref{fig2} and \ref{fig3}. If nanosized pores with a certain size were present, say $\sim$$10 \, \mathrm{nm}$, this would give rise to a spin-flip signal that is centered at around $\frac{2\pi}{10 \, \mathrm{nm}} \cong 0.6 \, \mathrm{nm}^{-1}$, which lies outside of the recorded momentum-transfer range.}.

From the neutron scattering point of view, this observation might be perceived as surprising since an isotropic ring structure may not be expected, given the highly anisotropic individual contributions to the spin-flip SANS cross section (compare Eq.~(\ref{eq:iucrj2023}) below or Eq.~(5) in \cite{nanoflower2026sm}). This finding supports the emergence of vortexlike magnetization correlations within the iron oxide nanoflower ensemble, since the ring structure can be qualitatively reproduced by the linear vortex model [Eq.~(\ref{eq:SpinFlip2DSANScrossSection})] using realistic values for the stiffness parameter $m_1 R / m_0$ and for the vortex-axis opening angle $\alpha_{\mathrm{c}}$ [see Fig.~\ref{fig2}(c)]. In fact, theory predicts that for a spherical vortex-state particle with a radius of $R$, and for the case that $m_1 R / m_0 \rightarrow  \infty$ (pure vortex) and $\alpha_{\mathrm{c}} \cong 68.5^{\circ}$, the intensity maximum of the ring is located at $q_{\mathrm{max}} = 2.5/R$~\cite{adamsprb2024no2,adamsprr2026}. Experimentally, we find $q_{\mathrm{max}} \cong 0.027 \, \mathrm{nm}^{-1}$ at $0.01 \, \mathrm{T}$, which corresponds to $R \cong 93 \, \mathrm{nm}$, close to the value extracted from electron-microscopy characterization [see Fig.~\ref{fig1}(a)].

To assess whether the isotropic spin-flip ring could already be explained by the zero-order, uniformly magnetized part of Eq.~(\ref{eq:LinearVortexModel}), i.e., without the finite-$m_1$ vortex contribution, we consider a more general dilute macrospin ensemble in which the orientational distribution is left arbitrary. Using the notation of Ref.~\cite{michaeliucrj2023}, the corresponding spin-flip SANS cross section reads:
\begin{align}
\label{eq:iucrj2023}
\frac{d\Sigma_{\mathrm{sf}}}{d\Omega} &\propto [f(qR)]^2 ( \Gamma_{xx} + \Gamma_{yy} \cos^4\theta \\
    &+ \Gamma_{zz} \sin^2\theta \cos^2\theta - 2 \Gamma_{yz} \sin\theta \cos^3\theta ) \nonumber ,
\end{align}
where the $\Gamma_{ij}$ are orientational second moments of the macrospin directions. A Fourier expansion of the angular terms shows that an isotropic spin-flip SANS cross section requires $\Gamma_{yy} = \Gamma_{zz} = \Gamma_{yz}=0$, leaving only the $\Gamma_{xx}$ contribution (see the Supplemental Material~\cite{nanoflower2026sm} for further details). This limiting case would correspond to macrospin directions aligned parallel or antiparallel to the neutron-beam direction, which is not a plausible configuration for the present field geometry. The observed isotropic ring therefore cannot be explained by a uniformly magnetized spherical-particle ensemble alone, but points to spatially nonuniform magnetic correlations, as captured by the finite-$m_1$ vortex term in Eq.~\eqref{eq:SpinFlip2DSANScrossSection}.

Figure~\ref{fig3} depicts the $2\pi$~azimuthally averaged spin-flip data $\langle I_{\mathrm{sf}}(q) \rangle$ at selected applied magnetic fields along with weighted nonlinear least-squares fits (solid lines) to the linear vortex model [Eqs.~(\ref{eq:SpinFlip1DSANScrossSection})$-$(\ref{eq:finalmodel1})]; the inset depicts the corresponding result for the 2D spin-flip SANS at $0.01 \, \mathrm{T}$. In this quantitative analysis ($W=1$), we have treated $\overline{R}, \sigma, \eta$ as global fit parameters, and the scaling factors $m_0$ and $m_1$ as local fit parameters (different at each field). Since we observed an isotropic ring type structure in our experiment, the angle $\alpha_{\mathrm{c}}$ (describing the vortex-axis distribution) was always fixed to a value of $68.5^{\circ}$~\footnote{The particular setting $\alpha_{\mathrm{c}} = 68.5^{\circ}$ results from the theoretical analysis~\cite{adamsprr2026} of the 2D spin-flip SANS cross section [Eq.~(\ref{eq:SpinFlip2DSANScrossSection})]. Due to the mathematical structure of the corresponding azimuthally averaged 1D expression [Eq.~(\ref{eq:SpinFlip1DSANScrossSection})], $\alpha_{\mathrm{c}}$ and $m_0$ cannot be independently determined from 1D spin-flip data alone, since the fit is sensitive only to their multiplicative combination.}. We obtain the following values: $\overline{R} = 102 \pm 1 \, \mathrm{nm}$, $\sigma = 1.97 \pm 0.05$, and $\eta = 0.21 \pm 0.01$. The ratio $m_1 \overline{R} / m_0$, which reflects the field dependence of the scattering, changes from $m_1 \overline{R} / m_0 \cong 103.5$ at $0.01 \, \mathrm{T}$ and $m_1 \overline{R} / m_0 \cong 5.8$ at $0.03 \, \mathrm{T}$ to $m_1 \overline{R} / m_0 \cong 3.3$ at $0.1 \, \mathrm{T}$. These effective values are compatible with the average flower size found by means of electron microscopy [Fig.~\ref{fig1}(a)] and with the expectation that the relative vortex contribution (related to $m_1 \overline{R} / m_0$) decreases with increasing field. The relatively large value for the effective broadening parameter $\sigma$ might be explained by contributions due to polydispersity, instrumental smearing, and unresolved interparticle or intraparticle correlations.

Micromagnetic simulations on single nanoflower microstructures provide further support for our neutron results; they demonstrate that vortex configurations form at low applied fields close to remanence once the nanoflower diameter exceeds about $70 \, \mathrm{nm}$~\cite{eliss2026,jefremovas2026nanoscale,nanoflower2026sm}. In these simulations, the dipolar energy supports the formation of vortex-type flux-closure patterns, where the individual, nearly single-domain cores (here, with a size of about $10 \, \mathrm{nm}$) act as the building blocks of the structure.


{\it Conclusion}---Using uniaxial polarization analysis, we have studied the magnetic microstructure of a $220 \, \mathrm{nm}$-sized iron oxide nanoflower multicore ensemble. The spin-flip scattering cross section has been analyzed using phenomenological expressions based on a linear vortex magnetization Ansatz [Eqs.~(\ref{eq:SpinFlip2DSANScrossSection})$-$(\ref{eq:finalmodel1})]. The experimental finding of an isotropic ring structure in the spin-flip SANS as well as its momentum-transfer dependence are well described by the model, adding further support for the presence of vortex-type spin correlations in this system at low applied magnetic fields. The isotropic spin-flip ring is also inconsistent with the zero-order uniform macrospin limit, rendering the finite-$m_1$~vortex interpretation more robust. Our results support a magnetic configuration in which the macrospins of the individual $10 \, \mathrm{nm}$-sized cores arrange in a dipolar-energy-driven flux-closure structure. The presented approach underscores polarized SANS as a powerful technique for the quantitative analysis of nonuniform magnetization states in bulk ensembles of magnetic nanoparticles, thereby complementing synchrotron-based magnetic transmission x-ray microscopy and surface-sensitive local probes (e.g., magnetic force microscopy or Lorentz transmission electron microscopy). Beyond the specific iron oxide nanoflower assemblies investigated here, the methodology may be extended to other systems hosting flux-closure and topologically nontrivial spin textures, including skyrmionic and hopfionic magnetic states.


{\it Acknowledgments}---M.P.A., \v{S}.L., and A.M.\ acknowledge financial support from the National Research Fund of Luxembourg (AFR Grant No.~15639149 and DeQuSky Grant No.~C22/MS/17415246). S.S., N.-J.S., and A.M.\ acknowledge financial support from the Marie Sk{\l}odowska-Curie COFUND Programme ``NEXTSTEP'' (Grant No.~101177133). E.M.J.\ acknowledges funding from the European Union's Horizon 2020 research and innovation program under the Marie Sk{\l}odowska-Curie Actions (Grant No.~101081455-YIA) and the Institute for Advanced Studies (IAS) of the University of Luxembourg. J.L.\ acknowledges the Ghent University special research fund (bof/baf/1y/2024/01/005 and bof/baf/1y/2025/01/017). M.P.M.\ acknowledges project PID2023-150170OB-I00. L.F.B.\ acknowledges project PID2023-146448OB. The authors thank the ILL for the provision of neutron beamtime.

{\it Data availability}---The data that support the findings of this article are not publicly available. The data are available from the authors upon reasonable request.


%

\end{document}